\documentclass[11pt, a4paper,
    DIV=15, 
    parskip=half,
    listof=numbered,
    headsepline=false, footsepline=false,
    headinclude, footinclude,
    open=any
]{scrreprt}
\usepackage[T1]{fontenc}
\usepackage[english]{babel}
\usepackage{microtype}
\usepackage{csquotes}

\usepackage[default]{lato}
 
\setkomafont{disposition}{\mdseries\rmfamily}

\newcommand{\traiberTitleFontSize}{35}
\newcommand{\traiberChapterFontSize}{18}
\newcommand{\traiberChapterFontParskip}{12}
\newcommand{\traiberSectionFontSize}{12}
\newcommand{\traiberSectionFontParskip}{10}
\newcommand{\traiberSubSectionFontSize}{10}
\newcommand{\traiberSubSectionFontParskip}{0}

\usepackage[onehalfspacing]{setspace}
\usepackage{xcolor}

\definecolor{traiberBlue}{rgb}{0.098, 0.675, 0.761}
\definecolor{traiberBlueDark}{rgb}{0.0, 0.467, 0.608}
\definecolor{traiberLight}{HTML}{D7EEF2}

\definecolor{traiberFontGrey}{rgb}{0.235, 0.235, 0.231}
\color{traiberFontGrey}
\AtBeginDocument{\colorlet{defaultcolor}{.}}

\usepackage[hidelinks]{hyperref}
\hypersetup{
  colorlinks=true,
  linkcolor=traiberBlue,
  urlcolor=traiberBlue,
  citecolor=traiberBlue
}
\usepackage{graphicx}
\usepackage{background}
\backgroundsetup{
  scale=1,
  angle=0,
  opacity=.75, 
  contents={\includegraphics[width=\paperwidth,height=\paperheight]{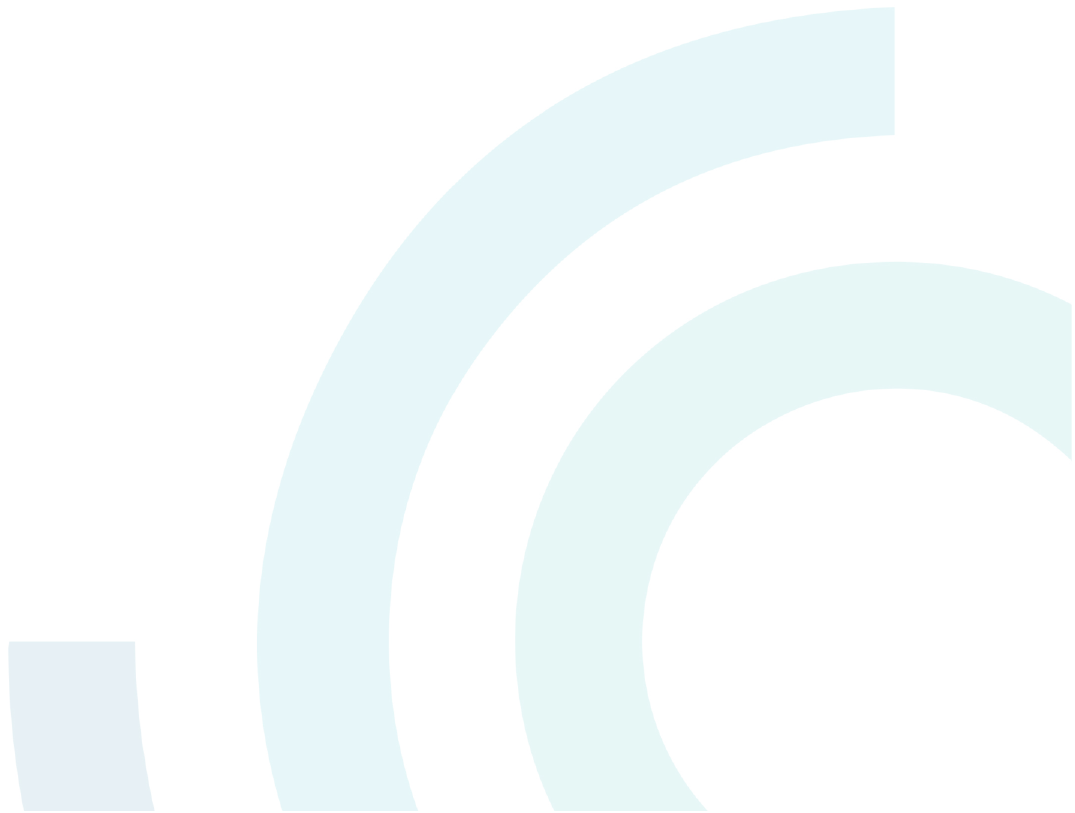}}
}

\usepackage{scrlayer-scrpage}
\clearpairofpagestyles
\ohead{\ifnum\value{page}<3\includegraphics[width=5.21cm]{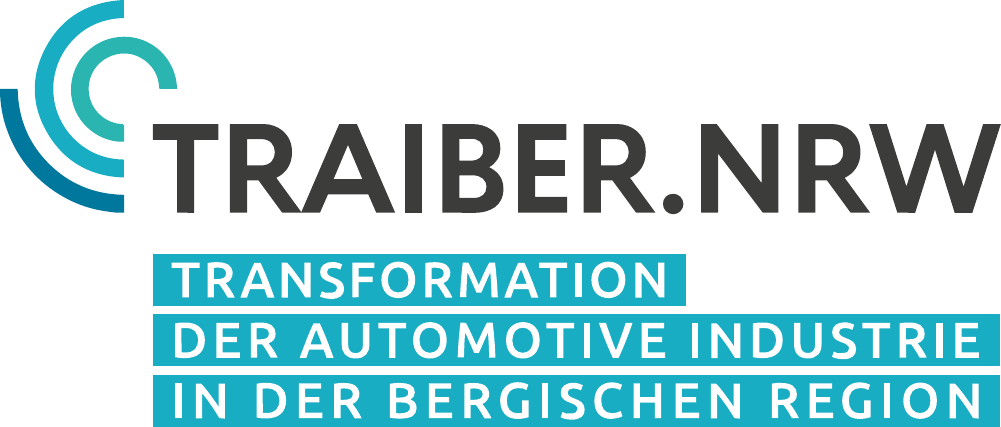}\else \includegraphics[width=5.21cm]{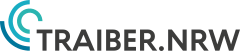}\fi}
\ihead{}
\ofoot{\ifnum\value{page}<3 \else \pagemark\fi}
\ifoot{}

\usepackage[top=4.2cm,bottom=2.85cm,left=2.9cm,right=2.49cm]{geometry}

\AddLayersToPageStyle{scrheadings}{bmwkfoot}

\RedeclareSectionCommands[
  beforeskip=1.\baselineskip,
  afterskip=.5\baselineskip
]{chapter}
\RedeclareSectionCommands[
  beforeskip=0.5\baselineskip,
  afterskip=.3\baselineskip,
  toclinefill=\hfill,  
]{section,subsection}

\setkomafont{chapter}{\fontsize{\traiberChapterFontSize}{\traiberChapterFontParskip}\selectfont\sffamily\color{traiberBlue}}
\setkomafont{section}{\fontsize{\traiberSectionFontSize}{\traiberSectionFontParskip}\selectfont\color{traiberBlue}}
\setkomafont{subsection}{\fontsize{\traiberSubSectionFontSize}{\traiberSubSectionFontParskip}\selectfont\color{traiberBlue}}
\setkomafont{descriptionlabel}{\color{traiberBlue}}

\addtokomafont{caption}{\small}
\setcapindent{1em}

\usepackage{enumitem}
\setlist[itemize]{nosep,left=1.2em}
\setlist[enumerate]{nosep,left=1.6em}

\usepackage[most]{tcolorbox}
\tcbset{
  enhanced,
  boxrule=0pt,
  colback=traiberLight,
  colframe=traiberLight,
  arc=2mm,
  left=6mm,right=6mm,top=4mm,bottom=4mm
}

\newcommand{\Doctitle}{Responsible AI in Business}
\newcommand{\Docsubtitle}{}

\newcommand{\ProjectLine}{Published as part of the TRAIBER.NRW project}
\newcommand{\ProjectInfoTitle}{PROJECT INFORMATION}
\newcommand{\ProjectInfoText}{%
The ongoing digitalization, the mobility transition, and the necessary shift towards climate neutrality are presenting numerous challenges for automotive suppliers and the Bergisches Land region. The Federal Ministry for Economic Affairs and Energy is funding TRAIBER.NRW with €4.1 million until the end of 2025 as part of the funding announcement "Transformation Strategies for Regions in the Automotive and Supplier Industry".

The Bergisches Land region comprises the cities of Remscheid, Solingen, Wuppertal, and Düsseldorf, as well as the districts of Mettmann, Rhein-Kreis Neuss, Ennepe-Ruhr, and Oberberg. Project partners include automotiveland.nrw e.V., the University of Wuppertal, the Joint Industrial Training Workshop of Velbert and Surrounding Areas (GLW), Heinrich Heine University Düsseldorf, and Bochum University of Applied Sciences. The project was initiated by the region’s social partners and is substantially supported and accompanied by them, including the Association of Bergisches Land Employers’ Associations ($\text{VBU}^{\textregistered{}}$), the Employers’ Association of Remscheid and the Bergisches Land e.V., IG Metall Velbert, IG Metall Ennepe-Ruhr-Wupper, and IG Metall Remscheid-Solingen.
}

\newcommand{\ProjectContact}{\url{www.traiber.nrw}}

\newcommand{\MakeTitlePage}{%
  \begin{titlepage}
    \BgThispage
    \thispagestyle{scrheadings}
    \vspace*{180pt}
    
    {\fontsize{\traiberTitleFontSize}{8}\selectfont\color{traiberBlueDark} \Doctitle\par}
    \vspace{0.6em}
    {\fontsize{20}{8}\selectfont\color{traiberBlueDark} \Docsubtitle\par}

    \vspace{310pt}
        
    \hspace{13.49cm} \includegraphics[width=3.67cm]{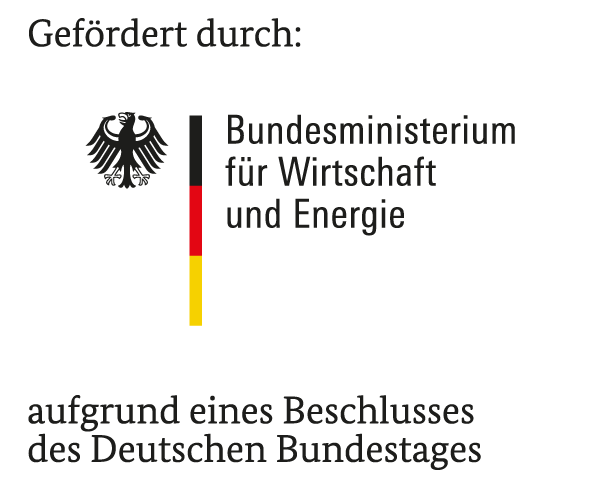}

  \end{titlepage}%
}

\newcommand{\MakeProjectInfoPage}{%
  \clearpage
  \BgThispage
  \thispagestyle{scrheadings}
  \null
  \vfill
  
  {\fontsize{20}{8}\selectfont\color{traiberBlueDark} \Doctitle\par}
  \vspace{0.25em}
  {\sffamily\itshape \ProjectLine\par}

  \vspace{1.6em}
  {\sffamily\bfseries\color{traiberBlue}\ProjectInfoTitle\par}

    \ProjectInfoText

    \vspace{0.25em}
    \ProjectContact
}

\usepackage[backend=biber,
style=ieee,
doi=false,isbn=false,url=true,maxbibnames=3]{biblatex}
\usepackage[toc, acronym]{glossaries}

\makeglossaries

\newglossaryentry{latex}
{
        name=latex,
        description={Is a mark up language specially suited for 
scientific documents}
}

\newglossaryentry{maths}
{
        name=mathematics,
        description={Mathematics is what mathematicians do}
}

\newglossaryentry{formula}
{
        name=formula,
        description={A mathematical expression}
}

\newacronym{gcd}{GCD}{Greatest Common Divisor}

\newacronym{lcm}{LCM}{Least Common Multiple}

\makeatletter
\newcommand\footnoteref[1]{\protected@xdef\@thefnmark{\ref{#1}}\@footnotemark}
\makeatother

\begin{document}
\pagestyle{scrheadings}

\MakeTitlePage

\MakeProjectInfoPage

\clearpage
\setcounter{page}{3}

{\hypersetup{hidelinks}
    \tableofcontents
}
\clearpage

\chapter{Responsible AI in a Business Context}
\label{sec:intro}

Artificial intelligence (AI) has made the leap from research and pilot projects into everyday business operations in recent years. The success of generative AI applications has made it clear that AI is no longer just a niche topic in computer science, but is broadly shaping business processes, products, and services.
Whether in quality assurance, production planning, customer service, or administration: AI-based systems support decision-making, automate processes, and unlock new business opportunities, especially where they are directly integrated into operational and industrial processes.
Current studies show that a large majority of German companies now consider AI business-critical and are increasing their budgets accordingly \autocite{kpmg2025press}. For organizations of all sizes, especially small and medium-sized enterprises (SMEs), the question is no longer whether AI will be used, but how this can be done responsibly and sustainably, i.e., in a legally compliant, explainable or comprehensible, resource-efficient and data-sovereign manner, without underestimating risks to security, quality, compliance and reputation.

This topic paper introduces the concept of ''Responsible AI'' and is aimed at decision-makers in organizations that use or plan to introduce AI, with a particular focus on small and medium-sized enterprises (SMEs).
The objective is to provide guidance on how AI can be introduced and further developed in a way that creates value, while keeping risks manageable and integrating AI into business processes as part of responsible digitalization moving beyond short term trends and with a view toward sustainable value creation.

The topic paper is structured to enable quick orientation. Technical details are only explored in greater depth where they are relevant for decision-making and implementation. For each topic area, a clear contextualization is provided, along with indications of which chapters are particularly important for which domains.
The main focus areas are:

\begin{itemize}
\item The \textbf{EU AI Act} as the legal framework for the use of AI in Europe,
\item \textbf{Explainable AI} as the basis for transparency and trust,
\item \textbf{Green AI} as an approach for resource-efficient and economically viable AI,
\item \textbf{Local models (on-premise/edge)} as the foundation for data-sovereign, controllable operation and strategic independence.

\end{itemize}

The position paper is structured in a modular way: each chapter begins with a brief contextual introduction suitable for all target groups and then elaborates on the aspects relevant to implementation and governance. For a \textbf{quick introduction}, the following overview assigns the chapters to typical areas of responsibility:

\begin{description}
  \item[\textbf{Chapter 2 -- EU AI Act:}]
  Executive management, legal/compliance, data protection. \\ 
  \textbf{Benefit:} Classification of obligations, roles, and risk categories. Basis for governance, documentation, and implementation requirements in the use of AI.

  \item[\textbf{Chapter 3 -- Explainable AI:}] Data science/analytics, IT, business unit leadership. \\ 
  \textbf{Benefit:} Guidance on which forms of explainability are required in practice. Bridge between model logic, technical comprehensibility, and trust in decisions.

\item[\textbf{Chapter 4 -- Green AI:}] ESG (Environmental, Social \& Governance), IT, controlling. \\
  \textbf{Benefit:} Leverage points for reducing energy and resource consumption. Connection between environmental impact, cost perspective, and technical implementation in AI operations.

  \item[\textbf{Chapter 5 -- Local Models (On-Premise/Edge):}] IT/architecture, information security \\ 
  \textbf{Benefit:} Decision-making basis for the operating model (local vs.\ cloud). Criteria for data sovereignty, security requirements, and infrastructure implications.

  \item[\textbf{Chapter 6 -- Summary \& Outlook:}] Executive management, legal/compliance. \\ 
  \textbf{Benefit:} Consolidation of the key messages. Structured guidance for next steps, priorities, and interfaces between areas of responsibility.
\end{description}

\chapter{The EU AI Act} \label{sec:aiact}

The \emph{EU AI Act} (Regulation (EU) 2024/1689) is the world’s first comprehensive law for the regulation of artificial intelligence. The AI Act follows a \emph{risk-based approach}: the obligations depend on how risky a particular use of AI is. The objective of the regulation is to harness the potential of AI responsibly without violating human rights, corporate rights, or applicable laws. International guidelines such as the OECD Principles on AI and the UNESCO Recommendation on the Ethics of Artificial Intelligence provide the normative foundation for this approach and emphasize a human-centered, transparent, and non-discriminatory design of AI systems. For small and medium-sized enterprises (SMEs), this creates new opportunities but also uncertainties and additional effort, as many tasks cannot be delegated to a legal department.
The use of AI thus becomes a management responsibility that brings together strategy and legal certainty, including potential liability issues. Depending on whether companies develop AI systems themselves, procure them, or merely use them, different obligations and evidentiary requirements apply, particularly for high-risk AI applications, for which the AI Act provides, among other things, risk assessments as well as documentation, transparency, and oversight obligations.

\section{Who is Covered by the AI Act?}

Private individuals who use AI solely for purely personal purposes are exempt from the EU AI Act. For companies, however, the following applies: \textbf{anyone who uses AI in a professional context falls under the scope of the law}. Even voluntary activities within an association can lead to the threshold of commercial use being exceeded.
In this context, the EU AI Act applies regardless of whether it has already been fully transposed into national law or not.

Central to the EU AI Act is a company’s own \emph{role}: Are you a provider of AI, for example as an operational tool, or are you a deployer?

\begin{itemize}
  \item \textbf{Provider:} Do you develop or modify an AI application and offer it under your own name or brand? In that case, you are a provider with extensive obligations. \emph{Example:} You use an open-source AI model such as Llama by Meta and build an internal tool that you make available to your team or to customers under your company name. Even if you “encapsulate” existing AI software into your own product and offer it, you are considered a provider.
  \item \textbf{Deployer:} Do you purchase or license AI software (e.\,g.\ a chatbot or automation software) and use it within your company without modifying the system yourself or marketing it under your own name? In that case, you are a deployer. \emph{Example:} You license an existing AI service for your website or for document classification. In this case, you fall under the deployer obligations.
\end{itemize}

A typical borderline case in practice is the \textbf{local operation of an AI model (self-hosting)}. If, for example, you operate a large language model (LLM) on your own servers and make it available to your employees, without making any substantial modifications and without distributing the system externally, you are generally also to be regarded as a provider. As long as you do not use high-risk AI, the regulatory effort usually remains manageable: no registration, no technical documentation obligations, but in particular transparency requirements (e.\,g.\ clearly stating that the interaction is with an AI system).

\section{Which AI Systems are Covered by the AI Act?}

Most of the AI Act revolves around the \emph{risk class} of the respective AI application. There are three main groups:
\begin{enumerate}
  \item \textbf{Prohibited practices:} AI systems that manipulate people or carry out biometric surveillance without consent are prohibited.
  \item \textbf{High-risk AI:} This category includes AI systems in sensitive areas such as medicine, the justice system, personnel recruitment, critical infrastructure, or education (e.\,g.\ automated performance assessment).
  \item \textbf{Other (low risk):} Most applications in everyday SME contexts, such as product recommendations in a webshop, chatbots for standard inquiries, spell checking, etc.\ fall into the category of \enquote{low risk}. For these, there are no registration obligations and no technical documentation requirements. Only transparency requirements apply, where applicable (notice when users are interacting with an AI).
\end{enumerate}

\section{What do Companies Specifically Need to do?}

The EU AI Act complements existing regulatory frameworks and applies \emph{without prejudice} to data protection requirements. As soon as AI systems process personal data, the GDPR remains fully applicable. This applies in particular to automated decisions with significant effects (Art.\,22 GDPR), as well as to transparency, purpose limitation, data minimization, and privacy by design.

There is currently movement in the system with regard to the classification of legal bases: with the Digital Omnibus, the European Commission proposes targeted clarifications on the application of legitimate interest (Art.\,6(1)(f) GDPR) in the context of AI systems and AI models, in particular along the AI lifecycle (e.\,g.\ training, testing, validation)\autocite{eucommission2025digitalomnibus}. Case law is pointing in a similar direction: in interim proceedings, the Higher Regional Court (OLG) of Cologne rejected an application for injunctive relief against the use of publicly available profile data for AI training and, upon summary review, generally classified the processing as capable of being based on Art.\,6(1)(f) GDPR\autocite{olgkoeln2025uki2_25}. When using external AI services, the \emph{type of contract} is decisive for data protection risks: for consumer offerings (e.\,g.\ ChatGPT Free/Plus), the use of content for model improvement can be prevented via opt-out, whereas business offerings (e.\,g.\ API, ChatGPT Enterprise/Business) are, according to provider statements, not used for training by default and support a data processing agreement (DPA)\autocite{openaiHowDataUsed,openaiEnterprisePrivacy}. For corporate use, contracts should therefore at least specify: roles and responsibilities (including Art.\,28 GDPR, where processing on behalf applies), no use for training without opt-in, rules on retention and deletion, as well as transparency regarding subcontractors and data transfers.

In addition to the GDPR, the EU AI Act introduces further, risk-based obligations. For SMEs, the first three points are typically always relevant. Additional requirements arise depending on the role (provider/deployer) and the risk class:

\begin{itemize}
    \item \textbf{Risk assessment and documentation:}
    First, it must be clarified: \emph{What role} do we have? Is our system high-risk? The delineation is not always trivial, many terms have not yet been conclusively clarified from a legal perspective. For high-risk systems, technical documentation, risk management, and, where applicable, registration with authorities are required.

    \item \textbf{Transparency obligations:}
    Users must be able to recognize when they are interacting with an AI (unless this is obvious anyway, e.\,g.\ an animated game character). Anyone who creates and publishes synthetic media (deepfakes, AI-generated images/videos/texts with potential to deceive) must label this, although there is flexibility regarding how visible such labeling must be. For artistic or satirical works, a general notice in context is sufficient.

    \item \textbf{AI literacy:}
    Providers and deployers of AI systems should, to the best of their ability, take measures to ensure a sufficient level of AI literacy among the persons who operate or use AI systems on their behalf. What constitutes “sufficient” is deliberately context and risk-based; the European Commission explicitly emphasizes flexibility in this regard and currently refrains from setting rigid minimum requirements. In practice, a baseline level of awareness is advisable for all employees who use AI-supported functions in their day-to-day work (e.\,g.\ Office Copilot, internal chatbots, video conferencing systems with automatic transcription or minute-taking). For higher-risk applications, especially high-risk systems, role-specific deepening is appropriate; the Commission also notes that additional requirements for training and human oversight may apply to high-risk systems.

    \item \textbf{General-purpose AI and open source:}
    Anyone who develops and offers their own AI models (in particular large, generative models) must provide additional transparency regarding training data, copyright strategy, and, where applicable, energy consumption. These obligations can generally only arise in the provider role. Open-source models benefit from certain simplifications, but specific obligations still apply here as well.
\end{itemize}

\section{Typical Challenges for Companies}

Even though many requirements of the EU AI Act are clearly formulated, additional clarification is often needed during implementation. Companies are confronted with legal as well as organizational and technical challenges. The following points are particularly relevant.
 
\begin{itemize}
  \item \textbf{Legal uncertainty:} Many definitions (e.\,g.\ when is a system considered high-risk?) will only become clearer through future practice and case law.
  \item \textbf{Documentation effort:} In the high-risk area, the effort is considerable. in the minimal-risk area, it usually remains manageable.
  \item \textbf{Liability:} Anyone who offers an AI system under their own name bears product responsibility. Simple rebranding solutions (API wrappers with one’s own logo) can quickly lead to being classified as a “provider” with all associated obligations. As a general rule, the company is (as a rule) fully liable for the use of AI, just as is the case in any commercial enterprise. The use of AI does not reduce liability in this context.
  \item \textbf{Technical implementation:} Labeling requirements (watermarks, metadata) cannot always be reliably implemented with existing tools. The state of the art is dynamic, and there are grey areas.
\end{itemize}

\section{Conclusion}
The AI Act is not an insurmountable obstacle, but it does require conscious decisions, careful clarification of roles, and, where appropriate, professional advice especially when using or offering high-risk AI.
Ultimately, a company remains liable for its products and their creation regardless of whether AI is used or not.
Those who use simple AI tools can plan with manageable effort and clear rules. Innovation is not prevented, but the scope for using AI depends heavily on how one’s own role and the specific purpose of use are defined.

\chapter{Explainable Artificial Intelligence}
\label{sec:xai}

Many modern AI models deliver impressively good results, but are hardly comprehensible from the outside. This creates a tension for companies: on one hand, decisions are meant to be efficiently automated, while on the other hand, they must be explainable to customers, employees, and supervisory authorities.
\textit{Explainable Artificial Intelligence} pursues the goal of making AI decisions more transparent and providing justifications that humans can understand. This chapter shows why explainability is not only a technical issue, but also a business-critical one.

Explainable AI therefore encompasses methods and techniques that make visible how an AI system arrives at a decision.
This is not only about technical details, but also about answering the question ''Can a human understand why the AI produces exactly this result?''.
XAI makes it possible to use AI without having to operate with an opaque black box \autocite{edps2023xai}, whose mechanisms and predictions are neither recognizable nor verifiable. One could also say that Explainable AI is an attempt to bridge the gap between incomprehensible machine decision-making processes and the human need for explanation and traceability. Three central concepts are distinguished:

\begin{itemize}
    \item \textbf{Transparency} refers to the comprehensible documentation and disclosure of relevant information about an AI system along its lifecycle. Depending on the context, this includes, among other things, the purpose and limitations of the system, performance metrics, as well as information on the origin and structure of the data used and on training, validation, and testing (e.\,g.\ data description, data preprocessing, evaluation setup) \autocite{EUAIActArt11,EUAIActArt13,ModelCards,Datasheets}.
    \item \textbf{Interpretability} means that a model is understandable to a human based on its structure, without requiring subsequently applied approximations or explanation methods. It should be directly comprehensible how a prediction comes about \autocite{MolnarIML,LiptonMythos}.
    \item \textbf{Explainability} describes methods that make the behavior or a specific decision of a model understandable for a particular purpose. This includes different forms (global vs.\ local) and techniques (e.\,g.\ attribution methods, counterfactuals, surrogates) \autocite{NISTIR8312,MolnarIML}.
\end{itemize}

Over the past ten years, Explainable AI has evolved from a research topic into a practically indispensable element in the use of AI in companies. There is hardly any way around explainable AI, if accountability, trust, and transparency are to be ensured. Wherever AI significantly influences decision-making processes, traceability is required. Transparency, in particular, is explicitly demanded by the EU AI Act (see Chapter \ref{sec:aiact}) and thus becomes an unavoidable topic for the use of AI systems. Explainability enables companies, beyond this, to open the \enquote{black-box AI} and thereby make AI more trustworthy and effective. This dovetails seamlessly with the principles of Responsible AI. Explainable AI creates transparency about how models function and helps to uncover bias and fairness issues before they lead to real harm. Bias and fairness issues arise when an AI system systematically treats certain groups differently from others due to faulty or imbalanced data.

\section{Why is Explainable AI Relevant for Companies?}

For companies, this topic is business-relevant for several reasons. AI is increasingly used in areas where decisions have direct economic consequences, such as quality control, resource planning, maintenance forecasting, or customer service. Without explainability, it remains unclear whether a model operates reliably, whether it exhibits systematic errors, or whether it produces decisions that violate internal guidelines or legal frameworks (see Chapter \ref{sec:aiact}). If this traceability is lacking, AI is quickly perceived as unsafe or risky, which slows down innovation, blurs responsibilities, and can significantly undermine acceptance of the use of AI. Anyone who has once seen an AI system make an error that cannot be explained loses trust.

With XAI, such risks can be controlled. Users and developers can examine whether a model relies on stable and plausible patterns, whether individual factors exert an excessively strong influence, or whether data distortions lead to unfair or inefficient outcomes. For companies, this means better decision-making foundations, lower liability risks, and greater trust in AI-supported processes. At the same time, Explainable AI facilitates communication between specialist departments and management, as explanations are typically presented in visually or linguistically easily understandable forms \autocite{Rong_2024}.

In corporate practice, the importance of Explainable AI becomes particularly clear when looking at concrete application scenarios. The following examples illustrate typical situations in which explainability directly determines trust, acceptance, and the quality of AI-supported decision-making. They can serve as guidance for identifying use cases within the company where XAI offers particular added value.

\begin{itemize}
  \item AI-based quality inspection, where it must be understandable why a component was classified as defective, in order to conduct complaints handling, audits, and internal quality discussions on a sound basis.
  \item Recommendation systems for products or services, whose suggestions should be explained in an understandable way for sales and customers, so that decisions are not adopted ''blindly''.
  \item AI-supported prioritization of service tickets or complaints, where employees must understand why certain cases are handled with priority, in order to ensure fairness and customer satisfaction.
  \item AI support for case processing, e.\,g.\ in claims handling, loan applications, or document review, where decisions must be justifiable because they have direct financial or legal consequences.
  \item AI-supported pre-selection of job applicants, where it must be transparent why certain candidates are preferred, in order to avoid discrimination and to be able to justify decisions to works councils and applicants.
  \item Internal reports and dashboards in which metrics or forecasts originate from AI models and must be explained to management and specialist departments in such a way that decisions can be based on them.
\end{itemize}

\section{Which Types of Explainable AI Exist in Practice?}
XAI encompasses different methods and perspectives. In practice, the following main categories are distinguished:

\begin{enumerate}
\item \textbf{Model-intrinsic explainability (intrinsic interpretability):}
These are models whose structure is inherently understandable. Their design makes it visible how individual inputs lead to a decision, for example through clear decision rules or manageable computational paths. Explainability thus arises directly from the model itself, without the need for additional analyses \autocite{doshivelez2017rigorousscienceinterpretablemachine}.

\item \textbf{Model-agnostic explainability (post hoc):}
In this approach, explanations are generated only after the actual model decision. This means that the model itself remains unchanged, but its behavior is analyzed retrospectively in order to provide understandable indications of influencing factors, decision logic, or possible alternatives. Such methods can, for example, show which features were particularly important for a given prediction or how the decision would change if individual inputs were different. Because these methods operate independently of the internal structure, they are also suitable for very complex, opaque models \autocite{Madsen2021PosthocIF}.

\item \textbf{Data-related explainability:}
Here, the focus is not on the model but on the dataset. The goal is to understand how the composition of the data influences model behavior. This includes, for example, identifying imbalances or missing groups, making potential biases visible, or tracing which data points have particularly strongly shaped the model. Such analyses help to assess how reliable and fair a model can operate \autocite{Suresh_2021}.

\item \textbf{LLM-specific explainability:}
Large language models not only generate outputs but often also produce accompanying justifications. These may sound plausible but are not always a reliable indication of the actual internal processes. Therefore, specialized approaches address making typical patterns within the model such as the processing of textual elements or the weighting of different signals more visible and comprehensible. The aim is to explain the functioning of such models in a more understandable way, without relying solely on the explanations formulated by the models themselves \autocite{zhao2023explainabilitylargelanguagemodels}.
\end{enumerate}

There are now mature approaches and tools for XAI that can be applied regardless of industry or company size \autocite{Yang2023SurveyOE,le2023p747}. They range from global model overviews that make the general functioning of a model visible, to local explanation methods that make individual decisions understandable in specific cases. This allows modern AI systems to be used not only effectively, but also in a verifiable, auditable, and strategically secure manner. In this way, AI evolves from a technology risk that is difficult to assess into a reliable tool.

\section{How can a Company Implement Explainable AI in Practice?}
Regulatory initiatives such as the EU AI Act (see Chapter \ref{sec:aiact}) explicitly emphasize the traceability of (high-risk) AI. At the same time, AI systems are becoming increasingly complex, this is particularly true in the context of generative AI which requires new explanation methods. Research is already working on even more advanced forms of explanation, ranging from counter examples and natural-language explanations to automated audit systems. For companies, it is important to develop an explainability strategy at an early stage: this includes selecting suitable XAI tools, training employees in the use of explainable AI, and embedding explainability in internal AI governance guidelines \autocite{vermeire2021chooseexplainabilitymethodmethodical}. First, it should be clarified: should the model \emph{only} perform well, or must decisions also be made understandable to users, supervisory authorities, or partners? Depending on the purpose, the required effort will vary.
To ensure that explainability does not remain merely an abstract principle but is actually implemented within the company, several fundamental measures are required. These relate both to the technical level of the models and to internal responsibilities and communication toward users \autocite{nguyen2024humancenteredexplainableaiinterface}.

The following three areas form the core of a practice-oriented approach to XAI implementation.

\vspace{-1em}
\paragraph{Ensuring Technical Explainability:}
Companies should implement at least basic explanation methods for each AI system:

\begin{itemize}
  \item global explanations (e.\,g.\ feature importance),
  \item local explanations (why exactly this decision?),
  \item data analyses (bias, outliers, representativeness).
\end{itemize}

\vspace{-1em}
\paragraph{Documentation and Responsibilities:}
Clear internal documentation is necessary so that responsibilities can be unambiguously assigned and all stakeholders can understand how an AI system is designed and operated. It forms the basis for auditability, quality assurance, and regulatory requirements. Relevant points include, among others:
\begin{itemize}
\item \emph{Which models are used?}
\item \emph{Which data form the basis?}
\item \emph{Which explanation methods are available?}
\item \emph{Who is responsible for updates and monitoring?}
\end{itemize}

\vspace{-1em}
\paragraph{User-Oriented Transparency:}
Explanations of the functioning and decisions of an AI system must be designed in a way that is understandable for the respective target groups, such as data analysts, customer service, management, or end users. Overly complex explanations are of no benefit.

\section{Conclusion}
Explainable AI is not an end in itself, but a tool for risk minimization, quality improvement and trust-building. For companies, XAI primarily means pragmatic orientation: simple methods, clear documentation, and understandable communication. Complex models remain complex, but with the right explanation methods, companies can make well-founded decisions, use AI safely, and deliver comprehensible results to both internal and external stakeholders.

In conclusion, Explainable AI is not merely an add-on, but an enabler for the successful use of AI in practice. It creates the foundation of trust on which humans and AI can jointly create value. Companies that design their AI systems to be explainable gain user and customer trust more quickly, save time during audits, and reduce liability risks. With explainability, a highly advanced AI solution ultimately becomes an accepted and reliable tool and thus represents a significant step toward truly responsible AI in the corporate world.

\chapter{Green AI}

\label{sec:greenai_sarah}
With the widespread introduction of large AI models, the ecological dimension of \textit{Responsible AI} is increasingly coming into focus.
High-performing AI models are often associated with high computational effort, energy consumption, and costs.
Companies therefore face the question of how the benefits of AI can be reconciled with economic and ecological objectives.
A central concept in this context is \textit{Green AI}: this refers to the requirement that AI systems should not only be evaluated based on their performance and accuracy, but also on their ecological footprint.
The focus is on energy and resource consumption (and the associated CO\textsubscript{2} emissions), particularly during the training and operation of very large models. Green AI emphasizes the efficient use of resources: instead of ''ever larger and more complex'', the emphasis is on reuse, optimization, and tailored solutions. This chapter shows how AI solutions can be designed in such a way that they remain not only performant, but also sustainable and economically viable.

Schwartz et al. showed in 2019 that the computational effort in deep learning increased by a factor of 300,000 between 2012 and 2018 \autocite{schwartz2019green}. Also in 2019, Strubell et al. quantified this effect exemplarily for NLP models (natural language processing models). They demonstrated that training individual large models can cause emissions in the range of several tonnes of CO\textsubscript{2} equivalents \autocite{strubell2019energy}.
In parallel, a regulatory framework is increasingly emerging: in addition to the expectations of investors in the area of ESG (Environmental, Social, and Governance), sustainability aspects, transparency requirements, and lifecycle considerations are also being discussed in the context of the EU AI Act (see Chapter \ref{sec:aiact}) \autocite{euaiact2024}.

\section{Green AI in Corporate Practice}\label{sec:Green_AI_Unternehmen}

Green AI is not only a technical issue, but directly influences cost structures, sustainability goals, and strategic decisions within companies.
The resource requirements of AI systems determine which solutions are economically viable and how well they can be aligned with environmental and ESG objectives.
In many cases, the ''largest possible'' model is not optimal; instead, a deliberately optimized, reused, or compressed variant is preferable.

Even if the benefits of Green AI cannot always be directly expressed as immediate financial added value for companies, they are strategically relevant from a long-term economic and ecological perspective. Companies that sufficiently take Green AI principles into account find it easier to meet regulatory requirements, reduce their costs in the medium to long term, and simultaneously strengthen their reputation.

Further efficiency gains arise when already trained internal models are deliberately reused or only incrementally retrained. This avoids additional training runs, thereby reducing computation time and energy consumption. Large models can be reduced in size using model compression methods, and the choice of more efficient algorithms as well as hardware tailored to actual needs can also be decisive factors in lowering resource requirements. In addition, systematic measurement of energy consumption contributes to greater transparency.

The following examples illustrate typical situations in which Green AI principles play a particularly important role and in which the approaches to model reuse and further training, model compression, and other measures described in the following sections can be applied concretely:

\begin{itemize}
  \item Introduction of a large language model for internal knowledge search, in which an existing language model is adapted to company data via transfer learning and LoRA fine-tuning instead of training a model from scratch. Through targeted model compression and an appropriate operating environment, both infrastructure costs and energy consumption are reduced.
  \item Use of AI for predictive maintenance, in which a model derived from a larger base model and subsequently compressed is executed directly on edge devices or industrial PCs, keeping computational effort, bandwidth usage, and energy consumption low during operation while simultaneously optimizing maintenance cycles.
  \item Development of an internal model suite in which a central base model is reused across multiple product lines and, for each use case, only small-scale adapters or finely tuned submodels are trained instead of building a completely new model for every task.
  \item Use of cloud resources for the (further) training of models, in which optimized training algorithms, suitable hardware profiles, and the deliberate limitation of training runs are used to actively manage both costs and the CO\textsubscript{2} footprint and make them measurable with the help of monitoring tools.
  \item Consideration of AI-specific consumption in ESG reporting and sustainability reports by systematically capturing the energy consumption and emissions of central models and transparently reporting the effects of model recycling, compression, and further efficiency measures.
\end{itemize}

\section{Reuse and Continuous Use of Models}\label{sec:GreenAI:WiederUndWeiterverwendung}
A particularly effective approach in corporate settings is the use of already pre-trained models instead of training models from scratch. This reduces development time, lowers infrastructure costs, and significantly decreases energy demand. Studies from various domains show that transfer learning achieves high performance with comparatively few additional training steps, while energy consumption is drastically reduced compared to full retraining \autocite{howard2018ulmfit, tschannen2023dora}. Practical guidelines, for example from the \emph{Green Software Patterns Catalog}, explicitly recommend evaluating pre-trained models as the default option in order to reduce the carbon footprint of AI projects \autocite{gsf2022patterns}. At the same time, new trade-offs may arise: centralizing on a small number of base or foundation models can increase dependencies on platform providers and raises questions regarding vendor lock-in, data sovereignty, and transparency of the underlying models. In principle, it must also be distinguished whether a model is used in a similar context or whether its context is to be expanded.
\label{subsec:tf_sarah}
\subsection{Transfer Learning}
The transfer of existing contextual knowledge to a similar application domain is referred to as \textit{transfer learning}. This is particularly advantageous for small datasets, where the amount of data is insufficient for full training. First, the model is trained on a larger, related dataset (e.\,g.\ standard German language), followed by fine-tuning with domain-specific data (e.\,g.\ technical terminology).
A modern and efficient technique for this is \textit{Low-Rank Adaptation (LoRA)}. In contrast to classical fine-tuning, only a subset of the model parameters typically individual layers is adapted, which enables fast and resource-efficient fine-tuning of large models. LoRA is based on the observation that weight matrices assume a low \textit{“intrinsic” dimension} during training, which allows for a simple low-rank decomposition. Studies show that LoRA reduces the number of trainable parameters for a GPT-3 model by a factor of 10,000 and lowers GPU memory requirements by a factor of three \autocite{hu2021loralowrankadaptationlarge}.
For companies, this means that existing models can be sensibly reused and adapted from both cost and environmental perspectives. A prerequisite is a sufficiently similar data basis in order to avoid negative transfer. Negative transfer refers to the performance degradation that occurs when a pre-trained model is applied to new data that differs too strongly from the original training data. In such cases, it may be necessary to train a model entirely from scratch.

\subsection{Continuous Learning}
When the model context is extended to include new use cases, this is referred to as incremental learning or \textit{continuous learning (CL)} (cf.\ \autocite{wang2024comprehensivesurveycontinuallearning}). Ideally, requirements are defined at an early stage so that suitable CL methods can be applied from the outset.
In the context of language models, CL refers to a model’s ability to continuously expand its knowledge by deliberately linking new information with existing knowledge, allowing the model to improve continuously and adapt to new language and content without forgetting what it has previously learned. A major challenge in this context is so-called \textit{catastrophic forgetting (CF)}, in which earlier learned content is overwritten as soon as new data is integrated \autocite{Kirkpatrick_2017}. This leads to a well-known dilemma \autocite{grossberg1980brain}: on the one hand, new tasks should be learned with sufficient flexibility, i.e.\ the model must be sufficiently plastic. On the other hand, existing knowledge should also be sufficiently preserved, which requires the model to maintain a certain degree of stability. Since these two aspects influence each other, an appropriate balance must be found. This is achieved through regularization methods, replay techniques, and architecture-related adaptations \autocite{wang2024comprehensivesurveycontinuallearning}.
Due to the high resource requirements of continuous training of large language models, efficient use of resources is essential, which can pose a challenge for smaller organizations. Smaller, specialized, or locally hosted models can be an alternative in this case, as can targeted model compression.

\section{Model Compression}
Model compression refers to a set of techniques that are deliberately applied to reduce the size of large, resource-intensive AI models in order to decrease their memory footprint and computational requirements while preserving performance as well as possible \autocite{neill2020overviewneuralnetworkcompression}. The goal is to make AI models efficiently usable even in resource-constrained environments such as mobile devices or edge computing, which also makes model compression a key enabling technology for the democratization of AI.
Important compression techniques include, among others:
\begin{itemize}
  \item \textbf{Pruning:} Removal of unnecessary connections or entire layers of a network \autocite{vadera2021methodspruningdeepneural},\autocite{10643325},\autocite{neill2020overviewneuralnetworkcompression}. In practice, pruning is particularly suitable for streamlining models for edge and predictive maintenance scenarios, enabling condition monitoring directly on industrial PCs or embedded systems with lower latency and reduced energy consumption.
  \item \textbf{Quantization:} Reduction of the numerical precision of model parameters (e.\,g.\ from 32-bit to 8-bit) \autocite{jacob2017quantizationtrainingneuralnetworks},\autocite{neill2020overviewneuralnetworkcompression}. Quantization is especially helpful where existing hardware is to continue being used, for example in internal language models for knowledge search: reduced numerical precision lowers memory and computational requirements without necessarily causing a noticeable loss in quality.
  \item \textbf{Knowledge distillation:} Transfer of knowledge from a large model to a smaller one \autocite{hinton2015distillingknowledgeneuralnetwork},\autocite{mansourian2025comprehensivesurveyknowledgedistillation},\autocite{neill2020overviewneuralnetworkcompression}. Knowledge distillation can be used to derive several smaller, task-specific models from a large, universal base model, which can then be efficiently operated within an internal model suite and deployed in a targeted manner for different departments or products.
  \item \textbf{Low-rank decomposition:} Decomposition of complex weight matrices into smaller components (cf.\ LoRA in Section~\ref{subsec:tf_sarah}) \autocite{hu2021loralowrankadaptationlarge},\autocite{neill2020overviewneuralnetworkcompression}.
  Low-rank decomposition and related methods such as LoRA are particularly useful when a shared base model is to be retained for many use cases, while adaptations are implemented in the form of lightweight adapters that can be trained quickly and deployed in a resource-efficient manner.
\end{itemize}

In corporate practice, these model compression techniques demonstrate their usefulness primarily by making powerful models manageable under real-world constraints.
By deliberately reducing model size, memory requirements, computational effort, and energy consumption can be significantly lowered without disproportionately impairing model quality.
This not only facilitates operation on existing infrastructure or resource-constrained devices (e.\,g.\ edge or industrial hardware), but also reduces ongoing costs and supports sustainability and ESG objectives.

At the same time, compressed models make it possible to use a single large base model in the form of smaller, specialized variants for different use cases.
Instead of training a new large model for each task, multiple lightweight models can be operated that build on a shared knowledge base and can be efficiently adapted.
Model compression thus represents a central building block for reconciling model reuse and continued use, economic viability, and ecological responsibility.

\section{Further Measures}
In addition to the reuse and continuous use of models as well as the targeted reduction of large models through model compression, there are further measures for more efficient use of resources. These primarily include the development and application of more efficient algorithms that can significantly reduce computational effort and, consequently, resource requirements. Such algorithms include, for example, adaptive and self-learning systems that dynamically adjust to changing conditions and thus use resources optimally. In addition, hardware optimizations play a key role, such as specialized chips that are explicitly developed for specific AI tasks and therefore operate in a particularly energy-efficient manner.
Another important aspect is the measurement and monitoring of the energy consumption of AI systems, which allows potential savings to be identified in a targeted way. All measures mentioned in this section can additionally contribute to the sustainable design of AI applications and can also be combined.

\section{Conclusion}
The integration of Responsible AI, and Green AI in particular, is indispensable for companies today. Current discussions about sharply rising emissions among major cloud providers such as the significant increase reported by Google due to growing AI workloads illustrate that Green AI is not a niche topic, but a central component of responsible corporate strategies \autocite{google2024report}. Sustainability strategies promote long-term competitiveness and strengthen corporate reputation. Recommended approaches from current guidelines aim to reduce environmental impacts such as CO\textsubscript{2} emissions and resource consumption while simultaneously enabling economic benefits. Instead of developing models from scratch, it should always be examined in advance whether existing models can be purposefully recycled and efficiently adapted through fine-tuning, specialized learning methods, dynamic architectures, and measures for model expansion or compression. Optimized algorithms and suitable hardware further contribute to energy savings. Responsible monitoring provides a transparent basis for assessing corporate and societal costs and strengthens the trust of all stakeholders. Ultimately, the balance between performance and resource use is decisive: while large models require more resources, smaller, modular, and compressed approaches are often sufficient. Strong centralization around a few base models also raises questions regarding dependencies, data sovereignty, and transparency.
For the successful integration of Green AI, companies should systematically track energy consumption and emissions, establish internal guidelines for model recycling, and always make decisions with a forward-looking view of accuracy, costs, environmental impact, and risks. Approaches such as LoRA offer pathways to modular, flexible, and resource-efficient model development. Green AI thus not only brings ecological improvements, but also enables cost savings, strengthens competitiveness, and secures the long-term viability of companies.

\chapter{Local Models (On-Prem/Edge) in the Responsible AI Context}\label{sec:lokaleModelle}

Before the widespread adoption of large cloud-based language models, the use of AI in companies was in many cases already a fully local matter, for example in the form of embedded systems in machines, vehicles, or production facilities.
With the availability of powerful generative models, however, the focus has shifted: many organizations initially test cloud-based LLMs, but then face the strategic question of which parts of their AI landscape should be operated in the cloud, on-premise, or directly on the ''edge'' in the long term. Especially for small and medium-sized enterprises (SMEs), this is not only about technical performance, but also about questions of data sovereignty, dependency on providers, and long-term operating costs.

In this chapter, ''local models'' are understood as AI systems that are executed entirely within an organization’s own infrastructure (on-premise) or directly on devices and machines (edge), in contrast to purely cloud-based AI services.
Local LLMs and edge AI make it possible to keep sensitive data within one’s own area of responsibility, make decisions with minimal latency, and adapt models deeply to company-specific requirements. At the same time, they bring their own challenges: investments in hardware, operational and maintenance effort, as well as the need to build internal expertise for the selection, operation, and securing of these systems.

In the Responsible AI context, local models are an important building block for bringing together several of the dimensions discussed above: they support data sovereignty and data protection (e.\,g.\ in light of the GDPR and the EU AI Act), enable domain-specific optimizations, and when combined with appropriate model sizes and architectures can contribute to Green AI by reducing computational effort and unnecessary data movement. At the same time, companies must weigh when local solutions actually offer advantages and in which scenarios cloud offerings remain sensible.


\section{Definitions of Local Models}

\textbf{Local AI models} are AI systems that are operated entirely within a company’s own infrastructure or directly on end devices (edge devices) in contrast to cloud-based AI services provided by external vendors. In an \textbf{on-premise model}, the AI model runs on the organization’s own servers or in its data center, meaning that all model data (weights, inputs, and outputs) remain within the corporate network \autocite{Zhang2025Jun}.
This gives the company full control over the model, including updates, configuration, and scaling, and ensures that sensitive data do not leave the internal network.
As a result, dependencies on external AI services such as vendor lock-in, outages, or sudden API price changes by third-party providers are eliminated \autocite{Dbm2025Nov}.
\textbf{Edge models}, by contrast, are integrated directly into decentralized devices or machines, for example in manufacturing facilities, vehicles, or IoT devices \autocite{singh2023edge}. They perform AI computations locally and in real time, without requiring a continuous connection to the cloud. This reduces latency, saves bandwidth, and increases resilience, as inference continues to function even when internet connectivity is limited or interrupted.

Compared to cloud-based AI services (where an external provider hosts the model in its cloud), local models therefore offer particular advantages in terms of data protection, control, and adaptability \autocite{trajanoski2025comparative}. Cloud-based LLMs, by contrast, are characterized by rapid availability and easy scalability. New AI functions can be integrated immediately via APIs, without the need to procure dedicated hardware. However, with cloud solutions, companies must accept that data leave their own secure network and are processed on external servers. The following comparison highlights the differences:
\begin{itemize}
    \item \textbf{Data control \& security}: Local LLMs offer full data sovereignty. All data processing takes place within a controlled internal environment, making them ideal for sensitive or regulated information. With cloud LLMs, responsibility for security largely lies with the provider; despite high cloud security standards, data are transmitted over the internet to data centers and are subject to corresponding access risks.
    \item \textbf{Adaptability \& integration}: On-premise models can be deeply customized, for example through fine-tuning with proprietary company data, custom tokenizers, or modifications to the model architecture. This flexibility allows domain knowledge and company-specific requirements to be directly embedded in the model. Cloud offerings, by contrast, usually provide only limited customization options (often restricted to provider-defined parameters or fine-tuning APIs).
    \item \textbf{Performance \& latency}: Locally operated AI models often respond with minimal latency, as no network transmission is required. This is particularly critical for real-time applications (e.\,g.\ in vehicles or production facilities). Cloud models, on the other hand, exhibit network-induced delays, and their performance depends on internet connectivity and the load on the provider’s servers.
    \item \textbf{Scalability \& operational effort}: Cloud services allow immediate scaling as needed (automatic addition of compute resources) and require lower initial investment. On-premise solutions offer consistent, dedicated compute capacity but must be expanded manually with additional hardware as demand grows. This entails higher upfront costs (for servers, GPUs, storage) and ongoing in-house maintenance effort. In return, for sustained intensive use, total costs can be more predictable and, in the long term, even lower than usage-based cloud fees.
\end{itemize}

\section{Data Sovereignty, Data Protection and Compliance}

A key advantage of local AI models is the preservation of data sovereignty. Companies retain full control over where and how their data are processed by AI systems. In contrast to cloud services, where input data leave the organization and are processed on external infrastructure, all data remain within the companys own area of responsibility when using locally operated models. This significantly facilitates the implementation of privacy by design: data protection is technically integrated from the outset. As a result, even highly sensitive information such as confidential development documents within a company or patient data in hospitals can be analyzed by AI systems without the risk of external data leakage \autocite{Makandra2025secureai}.
In practical terms, this means that when AI systems run on in-house servers or in domestic data centers, companies can ensure that all personal data remain protected in accordance with the GDPR. Legal grey areas associated with cloud-based AI usage (for example, whether providers located in third countries might gain access) can thus be avoided.
As a consequence, achieving \textbf{legal compliance} with respect to data protection becomes significantly easier especially in data-sensitive sectors such as healthcare, finance, and the public sector, where cloud solutions often reach regulatory limits.

\section{Domain-Specific Optimization: Fine-Tuning and RAG}

Beyond data protection considerations, local deployment also allows domain-specific knowledge to be optimally integrated. Companies can tailor their AI models precisely to their own use cases.
While this has long been standard practice for classical AI models, generic cloud-based LLMs are trained to be broadly applicable across many domains. Local LLMs, by contrast, can be enriched with industry-specific data, technical terminology, and proprietary documents.
Similarly, for edge AI applications, there may be more general or pre-trained models available, but in most cases these cannot be used directly and must be adapted to specific data, such as sensor readings or the objects to be classified. This places such adaptations in the realm of model reuse and continued use (see Chapter~\ref{sec:GreenAI:WiederUndWeiterverwendung}), which is why this section focuses primarily on generative AI and LLMs. For generative AI, there are two main approaches to domain-specific optimization: \textbf{fine-tuning} models with proprietary training data and \textbf{retrieval-augmented generation (RAG)} using internal corporate knowledge bases \autocite{yu2024evaluation}.

Through \textbf{fine-tuning} on internal text corpora, a language model can, for example, be adapted to legal terminology, medical reports, or technical jargon. The result is that the model produces more precise and relevant outputs within that specialized domain than a general-purpose model. Targeted fine-tuning of pre-trained language models can lead to substantial performance gains in specific text classification and generation tasks, a principle that is now widely applied in the context of LLMs.
On-premise models greatly simplify such adaptations: since the model weights are accessible, arbitrary additional training data can be incorporated, and even the architecture can be modified or specialized tokenizers used to optimally process company-specific terminology.
Cloud-based LLMs often do not offer this depth of customization; users are typically limited to the predefined functionalities of the provider. In a local environment, by contrast, AI teams can further train models using historical company data so that they ``understand'' proprietary product names, internal abbreviations, and internal guidelines. Multilingual adaptations are also possible: if an organization has extensive German-language knowledge, it can fine-tune a multilingual open-source model on German data or deploy a model already optimized for German to achieve better results in that language. The European AI startup ecosystem is producing such models such as Mistral AI from France, whose LLM variants are explicitly designed for European multilingualism and data protection.
These models run on European infrastructure and are trained on extensive German-, French-, and other European-language data, resulting in outputs that are more culturally and professionally appropriate.

Another key strength of on-premise models lies in their integration into existing IT ecosystems, allowing companies to connect their LLMs to internal data sources such as databases, knowledge management systems, or document repositories.
With \textbf{retrieval-augmented generation (RAG)}, a local model can query internal knowledge bases as needed in order to incorporate up-to-date facts or policies into its responses.
In this way, the LLM effectively becomes a company-specific knowledge repository that reflects internal terminology and practices.
All of this takes place within the organization’s protected environment, ensuring that all data remain on internal servers and that data sovereignty is preserved.
The advantage of RAG compared to fine-tuning is that the underlying knowledge base can be continuously expanded, whereas fine-tuning relies on a fixed snapshot of knowledge at the time of training.
When a company’s available knowledge grows, RAG only requires updating the databases rather than retraining the entire model.
This flexibility, however, also means that RAG offers less influence over the ``natural'' or learned capabilities of the model compared to fine-tuning.
RAG merely adds domain-specific context to each prompt, which can also increase the required internal network bandwidth.
Due to its high flexibility, however, RAG can be technically integrated into existing on-premise architectures with relative ease, allowing pilot projects to be rolled out more quickly.

The two approaches can naturally also be \textbf{combined}: stable or slowly changing core knowledge and stricter stylistic or linguistic constraints can be addressed through fine-tuning, while dynamic or extensible knowledge is retrieved via RAG from knowledge databases.
In summary, local models strengthen data sovereignty and enable AI systems to be tailored to industry- and company-specific requirements. Through privacy by design, sensitive information can be used securely, and through domain-specific adaptation these models often achieve higher accuracy and relevance for specialized tasks than generic AI services. This is a key factor for deploying AI responsibly and effectively in corporate contexts.

\section{Application Areas of Local Models}

Despite higher operational and maintenance effort, many companies choose local AI solutions when \textbf{data protection/data sovereignty}, \textbf{low latency}, or \textbf{domain-specific customization} are paramount. The following examples outline typical application areas for local LLMs (on-premise) and edge AI in corporate practice:

\begin{itemize}
    \item \textbf{Industry 4.0:} In manufacturing and production environments, edge AI models are primarily used for \emph{quality control} (e.\,g.\ visual inspection with cameras directly on the production line) and \emph{predictive maintenance}. Sensor and image data are analyzed locally to detect wear or anomalies at an early stage, without continuously transmitting large volumes of data to the cloud. This enables real-time responses, reduces bandwidth requirements, and increases robustness in the event of connectivity disruptions.

    \item \textbf{Automotive:} Modern vehicles integrate AI systems locally on board, for example for driver assistance systems and (partially) autonomous functions. Edge AI is essential here because safety-critical decisions must be made within milliseconds (e.\,g.\ object detection, lane keeping, emergency braking), and cloud latency is unacceptable. Local processing of camera and sensor data also increases resilience (e.\,g.\ in areas with poor connectivity) and protects privacy, as raw environmental data do not leave the vehicle.

    \item \textbf{Customer service \& internal assistant systems:} Many organizations deploy language-based assistant systems, such as chatbots for customer inquiries or internal knowledge assistants. With locally hosted LLM-based systems, confidential content (e.\,g.\ customer data, internal documents, protected process information) can be processed without passing data to third parties. This is particularly relevant in regulated sectors (e.\,g.\ finance, public administrati
    
\item \textbf{Healthcare:} In hospitals and telemedicine, locally operated AI models support, among other things, medical assistance systems (e.\,g.\ decision support for diagnosis and therapy) as well as patient monitoring via networked medical devices. Edge AI can evaluate vital signs in real time and trigger immediate alerts when anomalies occur. Processing data within hospital infrastructure facilitates compliance with data protection and confidentiality requirements (privacy by design), enabling AI applications to be operated in particularly sensitive contexts in a compliant manner.
\end{itemize}

In practice, a \textbf{hybrid approach} is often appropriate: cloud models can offer advantages for rapid prototyping and initial experimentation (scalability, low entry barriers). For productive, sensitive, or latency-critical use cases, smaller, fine-tuned models are then frequently deployed on-premise or at the edge. This allows innovation speed and control over data, operations, and compliance to be combined within an overall architecture.

\section{Contribution to Green AI: Energy Efficiency and Sustainability}

As already discussed in Chapter~\ref{sec:greenai_sarah}, not only computational performance but also the ``climate footprint'' is relevant when deploying AI models.
Edge AI approaches and small, specialized on-premise LLMs or multimodal models therefore represent important building blocks for greater sustainability. Instead of training a massive model from scratch for every task, the reuse of pre-trained models and their targeted fine-tuning can drastically reduce computational effort.
It is therefore advisable to evaluate pre-trained AI models as the default option in projects in order to reduce the carbon footprint. For companies, this results in a dual benefit: by leveraging existing base models (e.\,g.\ company-wide foundation models for text, image, or forecasting tasks), infrastructure costs can be reduced, time-to-market shortened, and emissions lowered simultaneously.
While the centralization on a small number of large models carries risks (such as dependency on individual providers and potential transparency issues in training), these can be mitigated through distilled or specialized smaller models. Such models are optimized for specific tasks or domains and can often achieve comparable performance with only a fraction of the computational resources. A conscious balancing act is required to select the smallest possible model that still delivers sufficient performance for each application, thereby avoiding unnecessary computation.

In addition, \textbf{edge AI} eliminates the need for permanent data transmission to centralized servers: when data are processed directly at the source (e.\,g.\ sensor data in industrial plants or video data in vehicles), this also saves the energy required to transmit, receive, and store large data streams in data centers. In the automotive industry, for example, local AI systems reduce the need to send vehicle data to the cloud and thereby \textbf{lower both costs and energy consumption for data transmission}, while enabling rapid diagnostics directly on site.
Furthermore, edge devices often allow for \textbf{optimized hardware acceleration}: specialized AI chips (e.\,g.\ NPUs in smartphones or FPGAs/ASICs in IoT devices) can perform inference with significantly lower power consumption per operation than general-purpose CPUs or GPUs in data centers. As a result, AI functions can be executed energy-efficiently ``at the edge'', which in aggregate can reduce overall electricity consumption, especially when many devices operate in parallel.

In summary, local and smaller AI models promote sustainability through \textbf{lower resource consumption}, avoidance of unnecessary data movement, and more efficient hardware utilization.
They enable companies to implement AI innovation while simultaneously minimizing the $\text{CO}_2$ footprint of their AI systems - an important criterion of responsible corporate governance.


\chapter{Summary \& Outlook}\label{sec:summary}

Responsible AI describes the aspiration to develop and deploy AI systems in such a way that they can be integrated into value creation in a \emph{legally compliant}, \emph{transparent}, \emph{sustainable}, and \emph{data-sovereign} manner. In everyday corporate practice, this means that AI is not merely an IT topic, but one that equally affects processes, products, responsibility, and reputation. The decisive shift in perspective is therefore no longer “Should we use AI?”, but rather “How do we use AI responsibly, in compliance with regulations, and in a way that is viable in the long term?”. Responsible AI brings together legal requirements, ethical guardrails, ecological objectives, and societal expectations into a manageable framework that both reduces risks and enables value creation: organizations that take a systematic approach lower liability and reputational risks, improve decision quality, and create the conditions for scaling AI reliably.

In the European context, the EU AI Act serves as a central point of orientation because it aligns obligations in a risk-based manner with the criticality of an application. For companies, this primarily means properly classifying AI use cases: which applications are prohibited, which are potentially high-risk, and which fall into less critical categories? Closely linked to this is the clarification of roles - whether a company acts as a \emph{provider} (developing, modifying, or placing AI on the market under its own brand) or as a \emph{deployer} (using purchased AI systems). This distinction is practically relevant because it determines the scope and depth of obligations, documentation, and responsibilities. From this classification follow typical implementation building blocks such as risk assessment and risk management, technical documentation, and, where applicable, registration. Additional requirements include transparency obligations. For example, when users interact with AI systems or when synthetic content has the potential to mislead as well as the establishment of \emph{AI literacy}: employees must be enabled, to an appropriate extent, to use AI correctly and to recognize risks.

Beyond regulation, data protection remains a core pillar, particularly in the context of automated decisions with significant effects. Key guiding principles here include robust legal bases, data minimization, purpose limitation, and privacy by design. In addition, corporate law requirements play a role, such as the protection of trade secrets and intellectual property. Especially when using external generative AI services, the risk of unintended disclosure of sensitive information or loss of control over data flows, models, and updates increases. Responsible AI therefore requires an integrated governance framework that brings together legal and technical risk assessments, bias and fairness checks, clear documentation and approval processes, human oversight, security and data protection measures, and continuous operational monitoring. AI is thus not introduced “somehow”, but rather as a controllable system with clear responsibilities throughout its entire lifecycle.

Another central lever is explainability, as many high-performing models are difficult to understand from the outside. Explainability addresses the tension between automation and responsibility by making visible how results are produced and which factors influence decisions. It is important to distinguish between \emph{transparency} (insight into processes, data, and documentation), \emph{interpretability} (models that are inherently understandable), and \emph{explainability} (methods that make decisions understandable for specific cases). In practice, depending on the context, this includes intrinsically interpretable model approaches, post-hoc explanations for complex models, and data-related analyses (e.\,g.\ bias, outliers, representativeness).
For companies, Explainable AI is business-critical because it increases trust and acceptance, facilitates communication between business units and management, and reduces liability and quality risks. Explainability, however, only becomes effective when minimum standards (global, local, data-related), clear responsibilities, and target-group-appropriate presentation are combined so that explanations are actually usable for management, specialist departments, and audit bodies.

With the widespread adoption of large models, resource consumption is also coming into sharper focus: computational effort, energy demand, and therefore costs increase, particularly during training and operation. Green AI therefore calls for AI to be evaluated not only based on performance metrics, but also on ecological impact and efficiency. For companies, this represents a dual opportunity, as efficiency measures reduce costs while supporting sustainability goals. Practical levers include reusing existing models instead of retraining, resource-efficient adaptations (transfer learning), incremental updates, model compression for operation, and monitoring and reporting of energy and resource usage. Often, a pragmatic approach is most effective: a smaller or adapted model may be sufficient for a specific process if it is robust, maintainable, and economically viable.

Finally, architectural decisions are a key component of Responsible AI, especially when data protection, control, latency, or resilience are critical. Local models (on-premise or edge) can strengthen data sovereignty and resilience because sensitive data do not need to leave the organization’s network and updates and configurations can be controlled. Edge approaches process data directly on devices or machines, reduce latency, save bandwidth, and remain functional even with poor connectivity. Cloud solutions, by contrast, offer rapid deployment and scalability, but require conscious management of data flows and dependencies. For domain adaptation, fine-tuning and retrieval-augmented generation (RAG) are particularly relevant: fine-tuning embeds company language and domain logic in the model, while RAG contextually integrates current internal knowledge from documents and knowledge bases. In practice, a hybrid approach is often appropriate, combining the exploratory and scalability advantages of the cloud with controlled operating models for sensitive, regulated, or production-critical scenarios.

In essence, Responsible AI emerges from the interplay of these dimensions: transparency requirements can enforce explainability, sustainability goals influence model and deployment decisions, and architectural choices directly affect data protection, security, and governance. For decision-makers, the key guiding principle is therefore not to optimize individual measures in isolation, but to establish a robust baseline assessment, clear responsibilities, and a step-by-step implementation plan that integrates roles, risks, evidence, and operational realities.

For decision-makers, the next sensible step is to derive a \emph{baseline assessment} from this classification: where is AI already being used, which risks and obligations arise, which evidence is missing, and which target states are realistic in the medium term?
The following list consolidates concrete, easily delegable work packages as \textbf{Next steps for your organization}.

\begin{itemize}
  \item \textbf{Build an AI landscape}: Create a central register of all AI applications (including shadow IT):
        purpose, process context, critical decisions, model/tool category, data sources, interfaces,
        external providers, and current maturity level (pilot, production, planned).

  \item \textbf{Classify role \& risk class}: Determine for each use case whether your organization primarily
        acts as a \emph{provider} or a \emph{deployer}, and assess the risk level of the application (e.\,g.\
        high-criticality decisions vs.\ supportive automation). Derive the key evidence, documentation,
        and control obligations from this classification.

  \item \textbf{Define governance \& responsibilities}: Establish clear responsibilities (e.\,g.\ RACI)
        for AI governance, legal/compliance, data protection, information security, business units,
        data science, and sustainability/ESG. Define decision points along the lifecycle (procurement,
        development, approval, operation, modification, decommissioning).

  \item \textbf{Define minimum transparency requirements}: Specify \emph{who} needs to understand what
        (management, business units, auditors, affected parties) and which explanation and evidence
        documents are required (e.\,g.\ purpose description, data provenance, model limitations,
        human oversight, test and acceptance protocols).

  \item \textbf{Secure operations}: Plan monitoring and incident processes: quality, drift, bias indicators,
        security incidents, logging, access control, update and change management, and supplier governance
        (contracts, SLAs, data flows, exit strategies).

  \item \textbf{Decide on sustainability \& architecture}: Measure or estimate resource consumption and costs
        of key AI workloads, assess reuse/compression/fine-tuning instead of retraining, and evaluate
        cloud vs.\ on-premise vs.\ edge across the dimensions of performance, data sovereignty,
        security, and energy efficiency.

  \item \textbf{Derive a roadmap}: Translate findings into a prioritized roadmap (quick wins in 4--8 weeks,
        medium-term capabilities in 3--6 months, strategic target states), including responsible owners,
        milestones, and clear success criteria.
\end{itemize}

\printbibliography[heading=bibintoc, title={References}]

@online{EUAIActArt11,
  author   = {{European Parliament and the Council of the European Union}},
  title    = {Regulation (EU) 2024/1689 (Artificial Intelligence Act), Article 11 (Technical documentation)},
  date     = {2024-06-13},
  note     = {OJ L, 2024/1689, 12.7.2024. See also Annex IV (technical documentation requirements for high-risk AI systems).},
  url      = {https://eur-lex.europa.eu/eli/reg/2024/1689/oj/eng},
  urldate  = {2026-01-10}
}

@online{EUAIActArt13,
  author   = {{European Parliament and the Council of the European Union}},
  title    = {Regulation (EU) 2024/1689 (Artificial Intelligence Act), Article 13 (Transparency and provision of information to deployers)},
  date     = {2024-06-13},
  note     = {OJ L, 2024/1689, 12.7.2024. Linked transparency obligations for high-risk AI systems.},
  url      = {https://eur-lex.europa.eu/eli/reg/2024/1689/oj/eng},
  urldate  = {2026-01-10}
}

@inproceedings{ModelCards,
  author    = {Mitchell, Margaret and Wu, Simone and Zaldivar, Andrew and Barnes, Parker and Vasserman, Lucy and Hutchinson, Ben and Spitzer, Elena and Raji, Inioluwa Deborah and Gebru, Timnit},
  title     = {Model Cards for Model Reporting},
  booktitle = {Proceedings of the Conference on Fairness, Accountability, and Transparency (FAT* '19)},
  publisher = {Association for Computing Machinery},
  address   = {New York, NY, USA},
  year      = {2019},
  pages     = {220--229},
  doi       = {10.1145/3287560.3287596},
  url       = {https://doi.org/10.1145/3287560.3287596},
  urldate   = {2026-01-10}
}

@article{Datasheets,
  author  = {Gebru, Timnit and Morgenstern, Jamie and Vecchione, Briana and Wortman Vaughan, Jennifer and Wallach, Hanna and Daum{\'e} III, Hal and Crawford, Kate},
  title   = {Datasheets for Datasets},
  journal = {Communications of the ACM},
  year    = {2021},
  volume  = {64},
  number  = {12},
  pages   = {86--92},
  doi     = {10.1145/3458723},
  url     = {https://doi.org/10.1145/3458723},
  urldate = {2026-01-10}
}

@book{MolnarIML,
  title     = {Interpretable Machine Learning},
  subtitle  = {A Guide for Making Black Box Models Explainable},
  author    = {Molnar, Christoph},
  year      = {2025},
  edition   = {3},
  isbn      = {978-3-911578-03-5},
  url       = {https://christophm.github.io/interpretable-ml-book},
  urldate   = {2026-01-10}
}

@article{LiptonMythos,
  author  = {Lipton, Zachary C.},
  title   = {The Mythos of Model Interpretability},
  journal = {Queue},
  year    = {2018},
  volume  = {16},
  number  = {3},
  pages   = {31--57},
  doi     = {10.1145/3236386.3241340},
  url     = {https://doi.org/10.1145/3236386.3241340},
  urldate = {2026-01-10}
}

@techreport{NISTIR8312,
  author      = {Phillips, P. Jonathon and Hahn, Carina A. and Fontana, Peter C. and Yates, Amy N. and Greene, Kristen K. and Broniatowski, David A. and Przybocki, Mark A.},
  title       = {Four Principles of Explainable Artificial Intelligence},
  institution = {National Institute of Standards and Technology},
  type        = {NIST Interagency/Internal Report (NISTIR)},
  number      = {8312},
  year        = {2021},
  doi         = {10.6028/NIST.IR.8312},
  url         = {https://doi.org/10.6028/NIST.IR.8312},
  urldate     = {2026-01-10}
}

@online{openaiHowDataUsed,
  author  = {{OpenAI}},
  title   = {How your data is used to improve model performance},
  year    = {2025},
  url     = {https://openai.com/policies/how-your-data-is-used-to-improve-model-performance/},
  urldate = {2026-01-10}
}

@online{openaiEnterprisePrivacy,
  author  = {{OpenAI}},
  title   = {Enterprise privacy at OpenAI},
  year    = {2025},
  url     = {https://openai.com/enterprise-privacy/},
  urldate = {2026-01-10}
}

@misc{eucommission2025digitalomnibus,
  author       = {{European Commission}},
  title        = {Proposal for a Regulation amending Regulations (EU) 2016/679, (EU) 2018/1724, (EU) 2018/1725, (EU) 2023/2854 and Directives 2002/58/EC, (EU) 2022/2555 and (EU) 2022/2557 as regards the simplification of the digital legislative framework (Digital Omnibus)},
  howpublished = {COM(2025) 837 final},
  year         = {2025},
  date         = {2025-11-19},
  url          = {https://eur-lex.europa.eu/legal-content/EN/TXT/?uri=CELEX:52025PC0837},
  urldate      = {2026-01-10}
}

@misc{olgkoeln2025uki2_25,
  author       = {{Oberlandesgericht K{\"o}ln}},
  title        = {Urteil vom 23.05.2025, 15~UKl~2/25 (KI-Training; Art.~6 Abs.~1 lit.~f DSGVO; Eilverfahren)},
  year         = {2025},
  date         = {2025-05-23},
  url          = {https://nrwe.justiz.nrw.de/olgs/koeln/j2025/15_UKl_2_25_Urteil_20250523.html},
  urldate      = {2026-01-10},
  note         = {ECLI:DE:OLGK:2025:0523.15UKL2.25.00}
}

@misc{kpmg2025press,
  author       = {{KPMG AG Wirtschaftspr{\"u}fungsgesellschaft}},
  title        = {Aus Kür wird Pflicht: 91 Prozent der deutschen Unternehmen sehen KI als geschäftskritisch an und stocken Budgets deutlich auf},
  howpublished = {Press release},
  year         = {2025},
  month        = {06},
  url          = {https://kpmg.com/de/de/home/media/press-releases/2025/06/aus-kuer-wird-pflicht-91-prozent-der-deutschen-unternehmen-sehen-ki-als-geschaeftskritisch-an-und-stocken-budgets-deutlich-auf.html},
  urldate      = {2025-09-05},
  note         = {Study “Generative KI in der deutschen Wirtschaft 2025”}
}

@article{schwartz2019green,
  title={Green AI},
  author={Schwartz, Roy and Dodge, Jesse and Smith, Noah A. and Etzioni, Oren},
  journal={Communications of the ACM},
  volume={63},
  number={12},
  pages={54--63},
  year={2019},
  doi={10.1145/3381831}
}

@article{Yang2023SurveyOE,
  title={Survey on Explainable AI: From Approaches, Limitations and Applications Aspects},
  author={Wenli Yang and Yuchen Wei and Hanyu Wei and Yanyu Chen and Guangtai Huang and Xiang Li and Renjie Li and Naimeng Yao and Xinyi Wang and Xiaodong Gu and Muhammad Bilal Amin and Byeong Kang},
  journal={Human-Centric Intelligent Systems},
  year={2023},
  volume={3},
  pages={161 - 188},
}

@inproceedings{le2023p747,
  title     = {Benchmarking eXplainable AI - A Survey on Available Toolkits and Open Challenges},
  author    = {Le, Phuong Quynh and Nauta, Meike and Nguyen, Van Bach and Pathak, Shreyasi and Schlötterer, Jörg and Seifert, Christin},
  booktitle = {Proceedings of the Thirty-Second International Joint Conference on
               Artificial Intelligence, {IJCAI-23}},
  publisher = {International Joint Conferences on Artificial Intelligence Organization},
  editor    = {Edith Elkind},
  pages     = {6665--6673},
  year      = {2023},
  month     = {8},
  note      = {Survey Track},
  doi       = {10.24963/ijcai.2023/747},
  url       = {https://doi.org/10.24963/ijcai.2023/747},
}

@article{Rong_2024,
   title={Towards Human-Centered Explainable AI: A Survey of User Studies for Model Explanations},
   volume={46},
   ISSN={1939-3539},
   url={http://dx.doi.org/10.1109/TPAMI.2023.3331846},
   DOI={10.1109/tpami.2023.3331846},
   number={4},
   journal={IEEE Transactions on Pattern Analysis and Machine Intelligence},
   publisher={Institute of Electrical and Electronics Engineers (IEEE)},
   author={Rong, Yao and Leemann, Tobias and Nguyen, Thai-Trang and Fiedler, Lisa and Qian, Peizhu and Unhelkar, Vaibhav and Seidel, Tina and Kasneci, Gjergji and Kasneci, Enkelejda},
   year={2024},
   month=apr, pages={2104–2122} 
}

@techreport{edps2023xai,
  title        = {Tech Dispatch on Explainable Artificial Intelligence},
  author       = {{European Data Protection Supervisor}},
  institution  = {EDPS},
  year         = {2023},
  month        = nov,
  url          = {https://www.edps.europa.eu/system/files/2023-11/23-11-16_techdispatch_xai_en.pdf},
  note         = {Accessed: 2025-02-14}
}

@misc{doshivelez2017rigorousscienceinterpretablemachine,
      title={Towards A Rigorous Science of Interpretable Machine Learning}, 
      author={Finale Doshi-Velez and Been Kim},
      year={2017},
      eprint={1702.08608}
}

@article{Madsen2021PosthocIF,
  title={Post-hoc Interpretability for Neural NLP: A Survey},
  author={Andreas Madsen and Siva Reddy and A. P. Sarath Chandar},
  journal={ACM Computing Surveys},
  year={2021},
  volume={55},
  pages={1 - 42},
}

@inproceedings{Suresh_2021, 
   series={EAAMO ’21},
   title={A Framework for Understanding Sources of Harm throughout the Machine Learning Life Cycle},
   url={http://dx.doi.org/10.1145/3465416.3483305},
   DOI={10.1145/3465416.3483305},
   booktitle={Equity and Access in Algorithms, Mechanisms, and Optimization},
   publisher={ACM},
   author={Suresh, Harini and Guttag, John},
   year={2021},
   month=oct, 
   pages={1–9},
   collection={EAAMO ’21} 
}

@misc{zhao2023explainabilitylargelanguagemodels,
      title={Explainability for Large Language Models: A Survey}, 
      author={Haiyan Zhao and Hanjie Chen and Fan Yang and Ninghao Liu and Huiqi Deng and Hengyi Cai and Shuaiqiang Wang and Dawei Yin and Mengnan Du},
      year={2023},
      eprint={2309.01029}
}

@misc{vermeire2021chooseexplainabilitymethodmethodical,
      title={How to choose an Explainability Method? Towards a Methodical Implementation of XAI in Practice}, 
      author={Tom Vermeire and Thibault Laugel and Xavier Renard and David Martens and Marcin Detyniecki},
      year={2021},
      eprint={2107.04427}
}

@misc{nguyen2024humancenteredexplainableaiinterface,
      title={How Human-Centered Explainable AI Interface Are Designed and Evaluated: A Systematic Survey}, 
      author={Thu Nguyen and Alessandro Canossa and Jichen Zhu},
      year={2024},
      eprint={2403.14496}
}

@misc{hu2021loralowrankadaptationlarge,
      title={LoRA: Low-Rank Adaptation of Large Language Models}, 
      author={Edward J. Hu and Yelong Shen and Phillip Wallis and Zeyuan Allen-Zhu and Yuanzhi Li and Shean Wang and Lu Wang and Weizhu Chen},
      year={2021},
      eprint={2106.09685}
}

@misc{wang2024comprehensivesurveycontinuallearning,
      title={A Comprehensive Survey of Continual Learning: Theory, Method and Application}, 
      author={Liyuan Wang and Xingxing Zhang and Hang Su and Jun Zhu},
      year={2024},
      eprint={2302.00487},
      archivePrefix={arXiv},
      primaryClass={cs.LG},
      url={https://arxiv.org/abs/2302.00487}, 
}

@article{Kirkpatrick_2017,
   title={Overcoming catastrophic forgetting in neural networks},
   volume={114},
   ISSN={1091-6490},
   url={http://dx.doi.org/10.1073/pnas.1611835114},
   DOI={10.1073/pnas.1611835114},
   number={13},
   journal={Proceedings of the National Academy of Sciences},
   publisher={Proceedings of the National Academy of Sciences},
   author={Kirkpatrick, James and Pascanu, Razvan and Rabinowitz, Neil and Veness, Joel and Desjardins, Guillaume and Rusu, Andrei A. and Milan, Kieran and Quan, John and Ramalho, Tiago and Grabska-Barwinska, Agnieszka and Hassabis, Demis and Clopath, Claudia and Kumaran, Dharshan and Hadsell, Raia},
   year={2017},
   month=mar, pages={3521–3526} }

@article{grossberg1980brain,
  title={How does a brain build a cognitive code?},
  author={Grossberg, Stephen},
  journal={Psychological Review},
  volume={87},
  number={1},
  pages={1--51},
  year={1980},
  publisher={American Psychological Association},
  doi={10.1037/0033-295X.87.1.1}
}

@misc{neill2020overviewneuralnetworkcompression,
      title={An Overview of Neural Network Compression}, 
      author={James O' Neill},
      year={2020},
      eprint={2006.03669},
      archivePrefix={arXiv},
      primaryClass={cs.LG},
      url={https://arxiv.org/abs/2006.03669}, 
}

@misc{vadera2021methodspruningdeepneural,
      title={Methods for Pruning Deep Neural Networks}, 
      author={Sunil Vadera and Salem Ameen},
      year={2021},
      eprint={2011.00241},
      archivePrefix={arXiv},
      primaryClass={cs.LG},
      url={https://arxiv.org/abs/2011.00241}, 
}

@article{10643325,
  author={Cheng, Hongrong and Zhang, Miao and Shi, Javen Qinfeng},
  journal={IEEE Transactions on Pattern Analysis and Machine Intelligence}, 
  title={A Survey on Deep Neural Network Pruning: Taxonomy, Comparison, Analysis, and Recommendations}, 
  year={2024},
  volume={46},
  number={12},
  pages={10558-10578},
  doi={10.1109/TPAMI.2024.3447085}
}

@misc{jacob2017quantizationtrainingneuralnetworks,
      title={Quantization and Training of Neural Networks for Efficient Integer-Arithmetic-Only Inference}, 
      author={Benoit Jacob and Skirmantas Kligys and Bo Chen and Menglong Zhu and Matthew Tang and Andrew Howard and Hartwig Adam and Dmitry Kalenichenko},
      year={2017},
      eprint={1712.05877}
}

@misc{hinton2015distillingknowledgeneuralnetwork,
      title={Distilling the Knowledge in a Neural Network}, 
      author={Geoffrey Hinton and Oriol Vinyals and Jeff Dean},
      year={2015},
      eprint={1503.02531}
}

@misc{mansourian2025comprehensivesurveyknowledgedistillation,
      title={A Comprehensive Survey on Knowledge Distillation}, 
      author={Amir M. Mansourian and Rozhan Ahmadi and Masoud Ghafouri and Amir Mohammad Babaei and Elaheh Badali Golezani and Zeynab Yasamani Ghamchi and Vida Ramezanian and Alireza Taherian and Kimia Dinashi and Amirali Miri and Shohreh Kasaei},
      year={2025},
      eprint={2503.12067},
      archivePrefix={arXiv},
      primaryClass={cs.CV},
      url={https://arxiv.org/abs/2503.12067}, 
}

@inproceedings{strubell2019energy,
  title={Energy and Policy Considerations for Deep Learning in NLP},
  author={Strubell, Emma and Ganesh, Ananya and McCallum, Andrew},
  booktitle={Proceedings of the 57th Annual Meeting of the Association for Computational Linguistics},
  pages={3645--3650},
  year={2019},
  doi={10.18653/v1/P19-1355}
}

@misc{euaiact2024,
  title={EU Artificial Intelligence Act},
  author={{European Parliament and Council of the European Union}},
  year={2024},
  note={Final legislative text}
}

@inproceedings{howard2018ulmfit,
  title={Universal Language Model Fine-tuning for Text Classification},
  author={Howard, Jeremy and Ruder, Sebastian},
  booktitle={Proceedings of the 56th Annual Meeting of the Association for Computational Linguistics (Volume 1: Long Papers)},
  pages={328--339},
  year={2018},
  doi={10.18653/v1/P18-1031}
}

@misc{tschannen2023dora,
  title={DORA: A Data-Efficient Approach for Fine-Tuning Foundation Models},
  author={Tschannen, Michael and Djolonga, Josip and Rubenstein, Paul K. and others},
  year={2023},
  archivePrefix={arXiv},
  eprint={2311.11829}
}

@misc{gsf2022patterns,
  title={Green Software Patterns Catalog},
  author={{Green Software Foundation}},
  year={2022},
  url={https://patterns.greensoftware.foundation}
}

@misc{google2024report,
  title={Environmental Report 2024},
  author={{Google Inc.}},
  year={2024},
  url={https://sustainability.google/reports/google-2024-environmental-report/}
}

@article{Zhang2025Jun,
	author = {Zhang, Zhoupeng and Shi, Jiaqi and Tang, Shaojie},
	title = {{Cloud or On-Premise? A Strategic View of Large Language Model Deployment}},
	year = {2025},
    journal = {{SSRN}},
	doi = {https://dx.doi.org/10.2139/ssrn.5296479},
}

@misc{Dbm2025Nov,
	author = {Database Mart},
	title = {{On-Premise vs Cloud LLM Hosting {\ifmmode---\else\textemdash\fi} Pros, Cons, and Use Cases}},
	journal = {Database Mart},
	year = {2025},
	month = nov,
	note = {[Online; accessed 19. Nov. 2025]},
	url = {https://www.databasemart.com/blog/on-premise-vs-cloud-llm-hosting}
}

@article{singh2023edge,
  title={Edge AI: a survey},
  author={Singh, Raghubir and Gill, Sukhpal Singh},
  journal={Internet of Things and Cyber-Physical Systems},
  volume={3},
  pages={71--92},
  year={2023},
  publisher={Elsevier}
}

@article{trajanoski2025comparative,
  title={Comparative Analysis of Large Language Models: On-Premise Architectures vs. Cloud-Based Deployments},
  author={Trajanoski, Stefan and Karadimce, Aleksandar},
  journal={Preface to Volume 5 Issue 2 of the Journal of University of Information Science and Technology “St. Paul the Apostle”--Ohrid},
  volume={5},
  number={2},
  pages={48},
  year={2025}
}

@misc{Makandra2025secureai,
	title = {{Local LLMs in organisations: Using AI securely}},
	author = {Makandra},
	year = {2025},
	note = {[Online; accessed 19. Nov. 2025]},
	url = {https://makandra.de/en/articles/local-llm-548}
}

@inproceedings{yu2024evaluation,
  title={Evaluation of retrieval-augmented generation: A survey},
  author={Yu, Hao and Gan, Aoran and Zhang, Kai and Tong, Shiwei and Liu, Qi and Liu, Zhaofeng},
  booktitle={CCF Conference on Big Data},
  pages={102--120},
  year={2024},
  organization={Springer}
}

\printglossary[title=Acronyms, type=\acronymtype]
\printglossary[title=Glossary]

\newpage

\null
\vfill

{\fontsize{20}{8}\selectfont\color{traiberBlueDark}{LEGAL NOTICE}}
\addcontentsline{toc}{chapter}{Legal Notice}
\vspace{2.0em}

\textcolor{traiberBlueDark}{PUBLISHER}
\vspace{1.25em}

\includegraphics[width=5.76cm]{Logos/Logo_TRAIBER/Traiber_Logo_WortBildMarke_RZ_RGB.png}
\vspace{1.25em}

\textcolor{traiberBlueDark}{AUTHORS \& EDITORIAL RESPONSIBILITY}

{\fontsize{10.65pt}{12pt}\selectfont Stephan~Sandfuchs\footnote{\label{traiberFootnote}TrAIBeR.NRW funded by the German Federal Ministry for Economic Affairs and Energy under grant reference 16TNW0024C}, Diako~Farooghi\footnoteref{traiberFootnote}, Janis~Mohr\footnote{CoFILL funded by the German Federal Ministry for Research, Technology and Space under grant reference 01IS24034B}, Sarah~Grewe\footnote{JetSki funded by the Ministry of Economic Affairs, Industry, Climate Action and Energy of the State of North Rhine-Westphalia under grant reference EFRE-20800516}, Markus~Lemmen\footnoteref{traiberFootnote} and Jörg~Frochte\footnoteref{traiberFootnote}
}\\
Interdisciplinary Institute for Applied AI and Data Science Ruhr (AKIS)\\
Bochum University of Applied Sciences

\vspace{2.5em}

\textcolor{traiberBlueDark}{TRAIBER.NRW ADMINISTRATIVE OFFICE}
\vspace{0.75em}

University of Wuppertal\\
Institute for Technologies and Management for Digital Transformation (TMDT)
\vspace{0.5em}

Building FZ | Level 01 | Room 19\\
Lise-Meitner-Str. 27--31, 42119 Wuppertal\\
Phone: +49 202 439-1164\\
Email: \url{koordination@traiber.nrw}\\
\url{www.traiber.nrw}
\vspace{0.25em}

Wuppertal, January 2026
\\




\end{document}